\documentstyle[preprint,prl,aps]{revtex} % for preprint form

\newcommand{\Tr}{{\rm Tr}\,}
\newcommand{\CN}{{\cal N}}
\newcommand{\CC}{{\cal C}}
\newcommand{\CM}{{\cal M}}
\newcommand{\CP}{{\cal P}}
\newcommand{\CF}{{\cal F}}
\newcommand{\CZ}{{\cal Z}}
\newcommand{\CO}{{\cal O}}
\newcommand{\CS}{{\cal S}}
\newcommand{\Pf}{{\rm Pf}}
\newcommand{\X}{{\bf X}}

\begin{document}
%\twocolumn
\draft
\title{\begin{flushright}
{\small\hfill AEI-058\\
\hfill hep-th/9803117}\\
\end{flushright}
Monte Carlo Approach to M-Theory}
\author{Werner Krauth \footnote{krauth@physique.ens.fr }}
\address{CNRS-Laboratoire de Physique Statistique,
Ecole Normale Sup\'{e}rieure\\
24, rue Lhomond\\ F-75231 Paris Cedex 05, France}
\author{Hermann Nicolai \footnote{nicolai@aei-potsdam.mpg.de}
$^{\S}$ {\it and} Matthias Staudacher \footnote{matthias@aei-potsdam.mpg.de }
\footnote{Supported in part by EU Contract FMRX-CT96-0012} }
\address{
Albert-Einstein-Institut, Max-Planck-Institut f\"{u}r
Gravitationsphysik\\ Schlaatzweg 1\\  D-14473 Potsdam, Germany }
%\date{\today}
\maketitle
\begin{abstract}
We discuss supersymmetric Yang-Mills theory dimensionally reduced
to zero dimensions and evaluate the $SU(2)$ and $SU(3)$ partition
functions by Monte Carlo methods. The exactly known $SU(2)$ results
are reproduced to very high precision.  Our calculations for $SU(3)$
agree closely with an extension of a conjecture due to Green
and Gutperle concerning the exact value of the $SU(N)$ partition
functions.

\end{abstract}
%\pacs{PACS numbers: 11.10.Kk, 11.10.St, 11.15.Pg, 11.15.-q, 12.38.Cy}
\vspace{5.5cm}
%{\small \hspace*{-1.7cm}AEI-xxx\\
%\hspace*{-1.7cm} 
%Supported in part by EU Contract FMRX-CT96-0012
%\begin{multicols}{2}
\newpage
\narrowtext

\section{The Model}

Recently there has been renewed interest in dimensionally reduced
supersymmetric Yang-Mills theories. It has been demonstrated that
reductions of the ten-dimensional theory to $p+1$ dimensions are
relevant to the description of $p$-dimensional extended objects
(D-branes) in string theory, and therefore, potentially, 
to quantum gravity. Furthermore, various such reductions are
currently being investigated in attempts to find a theory, termed
M-theory, which, it is hoped, non-perturbatively encompasses all known
superstring models as well as $D=11$ supergravity.
The most extreme such reduction, to zero dimensions, is
believed to describe string instantons, i.e.~configurations of open
superstrings whose ends are fixed in space-time.
The functional integral of Yang-Mills theory becomes an
ordinary multi-dimensional integral, which turns out to be
well-defined and {\it finite} if supersymmetry is present. 
The same integral appears
in attempts to rigorously define and calculate the Witten index of 
quantum-mechanical supersymmetric gauge theory \cite{Smilga,Sestern}. 
Lastly, it is the cornerstone of the so-called IKKT model~\cite{IKKT} 
which is claimed to yield a non-perturbative definition of superstring 
theory. 
Explicitly, the integral reads, for gauge group $SU(N)$, 
\begin{equation}
\CZ_{D,N}:=\int \prod_{A=1}^{N^2-1} 
\Bigg( \prod_{\mu=1}^{D} \frac{d X_{\mu}^{A}}{\sqrt{2 \pi}} \Bigg) 
\Bigg( \prod_{\alpha=1}^{\CN} d\Psi_{\alpha}^{A} \Bigg)
\exp \bigg[  \frac{1}{2} \Tr 
[X_\mu,X_\nu] [X_\mu,X_\nu] + 
\Tr \Psi_{\alpha} [ \Gamma_{\alpha \beta}^{\mu} X_{\mu},\Psi_{\beta}]
\bigg].
\label{susyint}
\end{equation}
We do not include a coupling constant into this definition,
since it could be immediately scaled out in a trivial fashion.
The matrices in the exponent in eq.(\ref{susyint}) are in the 
fundamental representation of $SU(N)$, i.e.
$X_{\mu}=X_{\mu}^A T_A$, $\Psi_{\alpha}=\Psi_{\alpha}^A T_A$, where
the $SU(N)$ generators $T_A$ are hermitean and normalized such that
$\Tr T^A T^B = \frac12 \delta^{AB}$. The symmetric $\CN\times \CN$ 
matrices $\Gamma^\mu$ are related to the standard $SO(1,D-1)$ gamma 
matrices by $\Gamma^\mu = \CC \gamma^\mu$, where $\CC$ is the charge 
conjugation matrix. The model is 
supersymmetric in dimensions $D=3,4,6,10$, where the degree
$\CN$ of (real) supersymmetry is, respectively, $\CN=2(D-2)=2,4,8,16$,
with the supersymmetry variations
\begin{equation}
\delta X_\mu = i \bar \varepsilon \gamma^\mu \Psi  \qquad
\delta \Psi = -\frac{i}{2} [X_\mu, X_\nu] \gamma^\mu \gamma^\nu \varepsilon.
\end{equation}
The most interesting case, in view of the above applications, is the
maximally supersymmetric integral with $D=10$.

We note that all calculations are performed with a Euclidean
signature. The Wick rotation to Euclidean space
is simply accomplished by the substitution
$X_0\rightarrow iX_{D}$. This leads to a positive definite bosonic 
``action'' in the exponent, which is a necessary (but not 
sufficient) prerequisite for the integral to be well-defined.
When switching from Minkowskian to a Euclidean 
signature, one in principle faces the problem of fermion
doubling because e.g.~the Majorana-Weyl spinors required
for maximally supersymmetric Yang-Mills theory exist only for
signature (1,9). However, this problem is spurious and completely
resolved by working only with the Euclidean 
Majorana field $\Psi$, and never making use of its complex 
conjugate field \cite{nic}. This is consistent because the relevant 
requirement in Euclidean quantum field theory is {\em not} 
hermiticity, but rather Osterwalder-Schrader reflection positivity.

Integrating out the real Grassmann variables $\Psi_{\alpha}^A$, one obtains
\begin{equation}
\CZ_{D,N} = \int \prod_{A=1}^{N^2-1} \prod_{\mu=1}^{D}  
\frac{d X_{\mu}^{A}}{\sqrt{2 \pi}}
\exp \bigg[  \frac{1}{2} \Tr
[X_\mu,X_\nu] [X_\mu,X_\nu] \bigg]~
\CP_{D,N}(X),
\label{int}
\end{equation}
where $\CP_{D,N}$ is a homogeneous polynomial of degree 
$k=(D-2) (N^2-1)=\frac12 \CN (N^2-1)$ in the variables $X_{\mu}^A$.
Its crucial property is invariance under $SO(D)\times SU(N)$. 
This polynomial is just the Pfaffian of the $2 k \times 2 k$ 
antisymmetric matrix
\begin{equation}
\big(\CM_{D,N}\big)_{\alpha \beta}^{A B}=-i f^{ABC} \Gamma^\mu_{\alpha \beta}
X_\mu^C,
\end{equation}
which can be directly read off from eq.(\ref{susyint}). 
Here $f^{ABC} = -2i \Tr T^A [T^B,T^C]$ are
the $SU(N)$ structure constants. The Pfaffian is 
explicitly given by
\begin{equation}
\CP_{D,N}=\Pf \CM_{D,N}=\frac{1}{2^k k!} \sum_{\sigma} (-1)^{\sigma}
\CM_{D,N}^{\sigma_1 \sigma_2} \ldots \CM_{D,N}^{\sigma_{2k-1} \sigma_{2k}}.
\label{pfaff}
\end{equation}
Here $\sigma$ runs over all permutations of the $2 k$ double 
indices $\sigma\equiv(\alpha A)$, and $(-1)^{\sigma}$ is the sign 
of the permutation. Although the matrices $\CM_{D,N}$ are complex in 
general, the Pfaffians are real as a consequence of the fact that we
started from a hermitean action (in Minkowski signature).
However, they are not necessarily positive. The explicit form of 
$\CM_{D,N}$ depends on the particular representation
of the Clifford algebra and the corresponding gamma matrices 
$\Gamma_{\mu}$. For definiteness, let us specify a convenient 
choice for the four dimensions of interest. 
Denote by $\bf{X}_{\mu}$ the adjoint representation of $SU(N)$,
i.e.~$\X_{\mu}^{AB}=f^{ABC} X_{\mu}^{C}$.

$\bullet D=3$: Here the Pfaffian is determined from the
$2(N^2-1) \times 2(N^2-1)$ matrix
\begin{equation}
\CM_{3,N}(X)=\left( \begin{array}{cc} 
\X_3 + i \X_2 & i \X_1  \\
i \X_1 & \X_3 - i \X_2
\end{array} \right) 
\end{equation}

$\bullet D=4$: In this case it is possible to reduce to a complex 
two-dimensional representation instead of the real four-dimensional 
representation. This allows us to rewrite the Pfaffian as an
ordinary determinant: there exists a
$2 (N^2-1) \times 2 (N^2-1)$ matrix $\CM'_{4,N}$ such that 
$\Pf \CM_{4,N}=\det \CM'_{4,N}$, with
\begin{equation}
\CM'_{4,N}(X)=\left( \begin{array}{cc} 
\X_4 + i \X_3 & i\X_2 + \X_1 \\
i \X_2 - \X_1 & \X_4 - i \X_3
\end{array} \right) 
\end{equation}

$\bullet D=6$: Again it is possible to replace the real 
eight-dimensional by a complex four-dimensional representation:
we have $\Pf \CM_{6,N}=\det \CM'_{6,N}$ with the 
$4 (N^2-1) \times 4 (N^2-1)$ matrix $\CM'_{6,N}$ 
\begin{equation}
\CM'_{6,N}(X)=\left( \begin{array}{cc} 
\X_6 + i \X_5 & i\X_4 + \sigma_1 \X_1 + \sigma_2 \X_2 + \sigma_3 \X_3  \\
i\X_4 -\sigma_1 \X_1 - \sigma_2 \X_2 - \sigma_3 \X_3 & \X_6 -  i \X_5
\end{array} \right) 
\end{equation}
where the $\sigma_i$ are the standard Pauli matrices.

$\bullet D=10$: In this case we use the decomposition 
\begin{equation}
\CM_{10,N}(X)=\left( \begin{array}{cc} 
\X_{10} + i \X_9 & i\X_8 + \Gamma^i \X_i  \\
i\X_8 - \Gamma^i \X_i & \X_{10} - i \X_9
\end{array} \right) 
\end{equation}
for the $16(N^2-1)\times 16(N^2-1)$ matrix $\CM_{10,N}$.
The indices $i,j,k$ run from 1 to 7, and the seven $8\times 8$ 
matrices $\Gamma^i$ satisfy the $SO(7)$ Clifford algebra
$\{ \Gamma^i,\Gamma^j\}= 2\delta^{ij}$.
A particularly nice 
representation is in terms of
the octonionic structure constants $c_{ijk}$ 
(again with $i,j,k=1,...,7$)
\begin{equation}
\begin{array}{ll}
 (\Gamma^i)_{jk} =& i c_{ijk} \qquad
 (\Gamma^i)_{88} = 0 \\
(\Gamma^i)_{j8} =& -(\Gamma^i)_{8j} = i \delta^i_j  
\end{array}
\end{equation}
These structure constants 
are completely antisymmetric; the non-vanishing ones are fully
specified by:
\begin{equation}
c_{124}=c_{235}=c_{346}=c_{457}=c_{561}=c_{672}=c_{713}=1
\end{equation}
The $\Gamma^i$ are obviously hermitean and purely imaginary.
Note that, unlike for $D=4$ and $D=6$, it is generally {\it not} 
possible here to reduce the Pfaffian to a determinant. 
The only exceptions are $N=2$ and $N=3$, since in this case the 
number of dimensions (=10) exceeds the dimension of the gauge
group, and we can use $SO(10)$ invariance to set $X_9=X_{10}=0$.
Then the Pfaffian becomes 
\begin{equation}
\CP_{10,N}(X)\big|_{X_9=X_{10}=0}= \det (i\X_8 + \Gamma^i \X_i)
\label{pfafftodet}
\end{equation}
This allows to integrate, for $SU(3)$ and $D=10$, 
over only eight instead of ten matrices, an observation that 
will prove useful in the numerical calculations below. After proper 
inclusion of the Fadeev-Popov determinant associated with
this reduction, the integral becomes
\begin{equation}
\CZ_{10,3}=\frac{1}{8!} \int \prod_{a,A=1}^8
\frac{d X_a^{A}}{\sqrt{2 \pi}}
\exp \bigg[  \frac{1}{2} {\rm Tr} \sum_{a,b=1}^{8}
[X_a,X_b] [X_a,X_b] \bigg]~
\Big[ {\rm det}_{1\leq c,C\leq 8} X^C_c \Big]^2
{\rm det} {\cal M}'(X)
\label{redint}
\end{equation}
where $\CM'(X)=i\X_8 + \Gamma^i \X_i$.

For general $D$ and $N$, the Pfaffian polynomials $\CP_{D,N}$ have a 
very complicated structure about which little is known. 
These complications are not yet fully apparent for $SU(2)$, where the 
Pfaffians are still comparatively simple \cite{Smilga,Sestern,Tsuch}:
\begin{eqnarray}
\CP_{3,2}(X)& = & \frac{1}{3} \epsilon^{ABC} \epsilon_{\mu\nu\rho} 
X_{\mu}^A X_{\nu}^B X_{\rho}^C =\\
 &=&
-\frac{2}{3} i
\epsilon_{\mu\nu\rho} \Tr X_{\mu} [ X_{\nu},X_{\rho} ] \\
\CP_{4,2}(X)& = & \Big. \frac{2}{3} \epsilon^{ABC} \epsilon^{A'B'C'}
X_{\mu}^A X_{\mu}^{A'} X_{\nu}^B X_{\nu}^{B'} X_{\rho}^C X_{\rho}^{C'}=\\
 & =&
\frac{8}{3} \Tr[X_{\mu},X_{\nu}] [X_{\nu},X_{\rho}] [X_{\rho},X_{\mu}] 
\Big. \\
\CP_{6,2}(X)& = & 
%\Big( \frac{2}{3} \epsilon^{ABC} \epsilon^{A'B'C'}
%X_{\mu}^A X_{\mu}^{A'} X_{\nu}^B X_{\nu}^{B'} X_{\rho}^C X_{\rho}^{C'} 
%\Big)^2 = 
\Big( \frac{8}{3} \Tr[X_{\mu},X_{\nu}] [X_{\nu},X_{\rho}]  
[X_{\rho},X_{\mu}] \Big)^2 \\
\CP_{10,2}(X)& = & 
%\Big( \frac{2}{3} \epsilon^{ABC} \epsilon^{A'B'C'}
%X_{\mu}^A X_{\mu}^{A'} X_{\nu}^B X_{\nu}^{B'} X_{\rho}^C X_{\rho}^{C'} 
%\Big)^4 = 
\Big(\frac{8}{3} \Tr[X_{\mu},X_{\nu}] [X_{\nu},X_{\rho}] 
[X_{\rho},X_{\mu}] \Big)^4 
\label{twopfaff}
\end{eqnarray}
Remember that the Greek indices in the preceding four equations are 
summed from $1$ to, respectively, $3,4,6,10$.
Already for $SU(3)$ and $D=3$, however, more ingenuity is required;
using symbolic computation, we have found the result (the 
$d^{ABC}$ are the $SU(3)$ symmetric structure constants)
\begin{eqnarray}
\CP_{3,3}(X)
=\Big( &\frac{3}{16} f^{ABE} d^{CDE} & f^{A'B'E'} d^{C'D'E'}-\cr
&-\frac{3}{10} f^{ABC} & d^{DA'E} f^{EB'E'} d^{E'C'D'} \Big)~ 
X_{\mu}^A X_{\mu}^{A'} X_{\nu}^B X_{\nu}^{B'} X_{\rho}^C X_{\rho}^{C'}
X_{\sigma}^D X_{\sigma}^{D'}
\label{fdfdone}
\end{eqnarray}
or equivalently,
\begin{eqnarray}
\CP_{3,3}(X)
=&-\frac{3}{4}~{\rm Tr}[X_{\mu},X_{\nu}] \{ X_{\rho},X_{\sigma} \}
~{\rm Tr}[X_{\mu},X_{\nu}]\{ X_{\rho},X_{\sigma} \}+\cr
 &~+\frac{6}{5}~{\rm Tr} X_{\mu} [X_{\nu},X_{\rho}]
~{\rm Tr} X_{\mu}\big[\{ X_{\nu},X_{\sigma} \},\{ X_{\rho},X_{\sigma} \}\big]
\label{fdfdtwo}
\end{eqnarray}

For higher dimensions and yet larger values of $N$, we expect the 
Pfaffians to increase rapidly in complexity. Let us nevertheless 
collect the following general relations whose validity follows upon
inspection of the explicit form of the matrices $\CM_{D,N}$ given 
above. Namely, we have, for any $N$,
\begin{equation}
\begin{array}{ll}
\CP_{10,N} (X)\big|_{X_7=...=X_{10}=0} &= 
\;\; \big(\CP_{6,N}(X)\big)^2    \\
\CP_{6,N} (X)\big|_{X_5=X_6=0} &= 
\;\;  \big(\CP_{4,N}(X)\big)^2    \\
\CP_{4,N} (X)\big|_{X_4=0} &= 
\;\;  \big(\CP_{3,N}(X)\big)^2   
\end{array}
\end{equation}
Of course, these relations are consistent with the explicit
$SU(2)$ results quoted above. We emphasize, however, that
relations of this type do not hold for generic configurations, 
as we have verified numerically. In addition, for $N\geq 3$ and 
$D \geq 6$, we have found configurations $X_\mu$ whose 
associated Paffians are {\it negative}. We also observe 
that $\CP_{D,N}(X)$ vanishes whenever the matrices $\X_\mu$ 
are restricted to lie in the regularly embedded $su(N-1)$ 
subalgebra of $su(N)$. Finally, $\CP_{D,N}(X)$ vanishes if 
only two $X_\mu$'s are different from zero.

Returning to the integral (3), the first question to ask is whether or
not it exists. The following naive argument seems to indicate that
$\CZ_{D,N}$ is divergent: Diagonalizing one of the matrices $X_{\mu}$
and restricting to the subspace where the off-diagonal elements 
of the other $D-1$ matrices are zero as well, we see that
the diagonal components of all $D$ matrices do not
appear in the action anymore. One might therefore suspect that the
fluctuations along the diagonals cause the integral to diverge.
Explicit calculations disprove 
this reasoning: it is shown in \cite{Sestern}
that for the gauge group $SU(2)$ the exact values
of the integral (\ref{int}) are\footnote{
While all the technical ingredients for this calculation can already be
found in \cite{Smilga}, the final result for the index obtained there
differs from the correct result in eq.(\ref{sutwo}) by a factor of $4$.
Also, that calculation did not take into account the deficit term of
\cite{Sestern} needed for the interpretation as a Witten index.}
\begin{equation}
\CZ_{D,2} =\sqrt{8 \pi} \times \left\{
\begin{array}{cc} 
0 & D=3 \\
&\\ 
\frac{1}{4} & D=4 \\
&\\ 
\frac{1}{4} & D=6 \\
&\\ 
\frac{5}{4} & D=10 
\end{array} \right.
\label{sutwo}
\end{equation} 
It is an instructive exercise to repeat the $SU(2)$ calculations 
for the case of {\it no} supersymmetry ($\CN=0$).
Omitting the Pfaffian, one obtains
\begin{equation}
\int \prod_{A=1}^{3} \prod_{\mu=1}^{D}  
\frac{d X_{\mu}^{A}}{\sqrt{2 \pi}}
\exp \bigg[  \frac{1}{2} \Tr
[X_\mu,X_\nu] [X_\mu,X_\nu] \bigg]=
\left\{
\begin{array}{cc}
\infty & \;\;\;D \leq 4 \\
&\\
2^{-\frac{3}{4}D -1} \frac{\Gamma(\frac{D}{4}) \Gamma(\frac{D-2}{4})
\Gamma(\frac{D-4}{4})}{\Gamma(\frac{D}{2}) \Gamma(\frac{D-1}{2})
\Gamma(\frac{D-2}{2})} & \;\;\;D \geq 5
\end{array}  
\right.
\label{nosusy}
\end{equation}
One concludes that despite the valleys of the potential, entropic 
effects of the measure can become strong enough to overwhelm
the possible divergences. Therefore we conjecture the integral 
$\CZ_{D,N}$ to be finite for all $N$, even though no rigorous 
proof is available for $N\geq3$ so far.
There are two sources of such entropic effects. 
One is the increase in dimension with larger $D$ and $N$: In the
above example, the integral becomes convergent if only a sufficient 
(here: five) number of random matrices is added. Furthermore, 
we generally expect the integrals to become better behaved
with larger $N$ since the relative weight of the valley 
configurations with regard to the generic configurations 
decreases like $N^{-1}$. Secondly, convergence is improved
by supersymmetry because the Pfaffian damps the contribution of the valleys:
It is easy to see that the Pfaffian vanishes if all $D$ matrices
commute simultaneously. 

Coming back to the supersymmetric case, we would like to find the
generalization of eq.(\ref{sutwo}) to higher rank gauge groups. 
Unfortunately the methods of \cite{Smilga,Sestern} are very special 
to $SU(2)$ and cannot be generalized to $N>2$ in any obvious way. 
On the other hand, exploiting the suspected
relationship of eq.(\ref{int}) to superstring instantons, Green and 
Gutperle \cite{greengut} made a bold conjecture as to the value of
the $D=10$ integral. Noticing, for $N=2$, a numerical coincidence between 
the result (\ref{sutwo}) and a term in their calculation of the 
D-instanton effective action, they suggested
\begin{equation}
\CZ_{D,N} \sim \sum_{m | N} \frac{1}{m^2} \;\;\;\; {\rm for}\;\;\;\;D=10
\label{conject}
\end{equation}
Here the sum runs over all divisors of $N$. If $N$ is prime,
the sum is simply $1+\frac{1}{N^2}$. Now it is clear that such
a conjecture is meaningless unless one very carefully specifies
all the {\it normalizations} in eq.(\ref{int})! 
Indeed, both the action and the polynomial $\CP_{D,N}$ are 
homogeneous functions of the $X_{\mu}^A$, and therefore (\ref{int}) is
a pure number up to rescalings: It is impossible to introduce
any non-trivial coupling constants. Fortunately, a natural normalization is
suggested by the index calculations of \cite{Sestern}. 
An $N$-dependent ``group factor'' $\CF_N$ has to be included, which
gives precisely the extra factor $\CF_2=\sqrt{8 \pi}$ in (\ref{sutwo}).
We can then extend the conjecture of \cite{greengut} to also include
the remaining dimensions $D=3,4,6$, and propose\footnote{
This proposal slightly differs, for $D=4,6$  and non-prime values of $N$, 
from an earlier version of the present article. We thank I.K.~Kostov
for alerting us to the fact that the present form is indeed more natural.}

\begin{equation}
\CZ_{D,N}=\CF_N \times \left\{ 
\begin{array}{cc}
0  & D=3 \\
&\\ 
\frac{1}{N^2} & D=4 \\ 
&\\
\frac{1}{N^2} & D=6 \\
&\\ 
\sum_{m | N} \frac{1}{m^2} & D=10
\end{array} \right.
\label{ourconject}
\end{equation}
We note that the result $\CZ_{3,N}=0$ for {\em even}
$N$ is a trivial consequence of the reflection antisymmetry
$X_\mu\rightarrow -X_\mu$; on the other hand, as our formulas 
(\ref{fdfdone}),(\ref{fdfdtwo})
for the $D=3,N=3$ Pfaffian shows, the result (if true!) is all
but trivial for odd $N$.

It seems natural to assume that $\CF_N$ is {\it independent} of
the dimension. Extending the assumptions going into the
index calculation of \cite{Sestern} 
to arbitrary $N \geq 2$ we find, with our
normalizations of the integral (\ref{int}),
\begin{equation}
\CF_N=\frac
{2^{\frac{N(N+1)}{2}} \pi^{\frac{N-1}{2}}}
{2 \sqrt{N} \prod_{i=1}^{N-1} i!}
\label{groupfactor}
\end{equation}

It should be stressed that our extended conjecture (\ref{ourconject}),
if true, furnishes an {\it exact} analytic solution of the IKKT matrix model
\cite{IKKT}. Indeed, the partition function of the IKKT model is defined
by summing up the $D=10$ $SU(N)$ partition funtions, and we thus propose
\begin{equation}
\CZ_{\rm IKKT}(\beta)=\sum_{N=0}^{\infty}~\CZ_{10,N}~e^{-\beta N}
\end{equation}
with $\CZ_{10,N}$ given as in eqs.(\ref{ourconject}),(\ref{groupfactor}),
and $\CZ_{10,0}=\CZ_{10,1}=1$. The weakest part in our proposal is
the form (\ref{groupfactor}) of $\CF_N$ (see also our results below):
Firstly, we might still have missed an $N$-dependent factor
in eq.(\ref{groupfactor}), and secondly,
the correct definition of the sum over $N$ might involve a yet
unknown $N$-dependent constant ${\cal C}_N$:
\begin{equation}
\CZ_{\rm IKKT}(\beta)=\sum_{N=0}^{\infty}~{\cal C}_N~
\sum_{m | N} \frac{1}{m^2}~
~e^{-\beta N}\
\label{IKKTsol}
\end{equation}

It is interesting to observe that for $D=10$ the proposed form 
(\ref{ourconject}) of $\CZ_{D,N}$ is {\it non-analytic} in $N$:
For {\it any} real number $c$ with $1 \leq c \leq \frac{\pi^2}{6}$ we
can find a sequence $\{N_\omega\}\subset {\rm I}\!{\rm N}$ such that
$\lim_{\omega\rightarrow\infty}\sum_{m | N_\omega} \frac{1}{m^2} = c$.
Such a result, if true, would give an explicit example of a case
where even the  assumption of the existence of a large $N$ limit is
simply wrong. On the other hand, it is intriguing for the future
that summing over all values of $N$, as proposed for the IKKT model,
clearly reinstates analyticity (as long as ${\cal C}_N$ is analytic), 
since we can rewrite eq.({\ref{IKKTsol}) quite generally as
\begin{equation}
\CZ_{\rm IKKT}(\beta)=\sum_{m=1}^{\infty}~\frac{1}{m^2}~\sum_{k=1}^{\infty}~
{\cal C}_{m k}~e^{-\beta m k}
\end{equation} 
 
Clearly these conjectures are very much in 
need of direct evidence from the actual integral (\ref{int}).
Analytical methods to treat this type of integral for $N>2$ are not
available for the moment, and we have therefore concentrated on a
numerical evaluation of the $SU(2)$ and $SU(3)$ partition functions using 
Monte Carlo methods, which, as we will show, need to be quite
intricate in order to cope with the large variations of  the integrand, 
provoked by the presence of the valleys.

\newpage

\section{Monte Carlo Approach}  

In the following, to simplify notations, we write the integral 
eq.(\ref{int}) as
\begin{equation}
\CZ_{D,N} = \int_V dx~z_{D,N}(x)
\end{equation}
where $d=D(N^2 -1)$ is the total dimension of the integral,
with  $x =(x_1, x_2, \ldots,x_d)$.
The  volume $V$, for the moment, is $V= {\rm I}\!{\rm R}^d$.
For a Monte Carlo evaluation of widely varying  integrals (cf \cite{recipes}), 
we generally
attempt to split off a positive {\em importance sampling} function
$\pi(x)$: $z_{D,N}(x) = \pi(x) \times \tilde{z}_{D,N}(x)$ such that
\begin{equation}
\int_V dx~z_{D,N}(x) = \int_V dx \pi(x) \;\; \tilde{z}_{D,N}(x).
\end{equation}
A judicious choice of $\pi$ 
will permit us to preferentially generate points in the valleys. 
This will much reduce the statistical fluctuations:
Suppose we are able to draw $M$ independent random configurations
$x_i$, $i=1, \ldots, M$ distributed according to the probability
$p(x) \sim \pi(x)$ with $V_{\pi}=  \int_V dx \pi(x)$ . We then
have
\begin{equation}
\CZ_{D,N} = V_{\pi} \times \frac{1}{M} \sum_{i=1}^{M} \tilde{z}_{D,N}(x_i) +
\CO(\frac{\sqrt{ \delta^2 \CZ_{D,N} }}{\sqrt{M}})
\label{direct}
\end{equation} 
with
\begin{equation}
\delta^2 \CZ_{D,N} = V_{\pi}^2 \times \frac{1}{M} \sum_{i=1}^{M}
(\tilde{z}_{D,N}(x_i))^2 - V_{\pi}^2 \times (\frac{1}{M} \sum_{i=1}^{M}
\tilde{z}_{D,N}(x _i))^2
\end{equation}
The unavoidable statistical error will therefore depend on the second moment
\begin{equation}
\int_V dx \pi(x) \;\; \tilde{z}_{D,N}^2(x)
\label{variance}
\end{equation}
In principle, the Monte Carlo 
computation is possible only with choices of $\pi$ for which 
the second moment (\ref{variance}) exists.
It should also be realized that, as is evident from eq.(\ref{direct}),
we need to know $ V_{\pi}$, the integral of $\pi$. If the latter
is unknown, we are usually only able to evaluate expressions of the type
\begin{equation}
\frac{\CZ_{D,N}}{ V_{\pi}} = 
\frac{\int dx \pi(x)~\tilde{z}_{D,N}(x)}{\int dx \pi(x)}
\end{equation}
We summarize these last remarks by stating that every approximation
$\pi(x)$ to the integrand  $z_{D,N}$ in  eq. (\ref{int}) can serve
as a useful starting point for very precise Monte Carlo evaluations
of the integral - for arbitrarily high  dimension $d$ -  
provided that $\int_V dx \pi(x) \;
\tilde{z}_{D,N}^2(x)$ exists, that $|z_{D,N}(x)| \sim \pi(x)$
everywhere and, lastly, that the integral of $\pi(x)$ be known.

As a first step towards the evaluation of $\CZ_{D,N}$, we {\em
compactify} $z_{D,N}$ onto the surface of the $d$-dimensional
unit hypersphere $S^d$. To do this, we introduce polar coordinates,
which we succinctly denote by $(x_1, \ldots, x_d) = (\Omega_d, R)$.
The action $\CS$ 
\begin{equation}
{\cal S}=-\frac{1}{2} {\rm Tr} \sum_{\mu=1,\nu=1}^{D}
[X_\mu,X_\nu] [X_\mu,X_\nu]
\end{equation}
is homogeneous in $x$:
\begin{equation}
\CS(\Omega_d,R) = \CS(\Omega_d,1) \times R^4
\end{equation}
as is the Pfaffian ($k=(D-2) (N^2-1)$)
\begin{equation}
\CP_{D,N}(\Omega_d,R) = \CP_{D,N}(\Omega_d,1) \times R^k
\end{equation}

The $R-$integration can now be performed exactly, and we find
\begin{equation}
\CZ_{D,N} = \frac { \int {\cal D}\Omega_d~z_{D,N}(\Omega_d) }
{ \int {\cal D}\Omega_d  }
\label{havelz1}
\end{equation}
with
\begin{equation}
z_{D,N}(\Omega_d)=
2^{(N^2-1) \frac{D-4}{2} -1}
\frac{ \Gamma\Big((N^2-1) \frac{D-1}{2}\Big) } 
{ \Gamma\Big((N^2-1) \frac{D}{2}\Big) } \times 
\frac{\CP_{D,N}(\Omega_d,1)}
{\Big[ \CS(\Omega_d,1) \Big]^{\frac{D-1}{2} (N^2-1)}}
\label{havelz2}
\end{equation} 
Without being obligatory, the compactification has 
nevertheless considerably smoothed the integrand, whose $R$-dependence 
was extremely strong.
Eq.(\ref{havelz2}) means that we can evaluate the integral eq.(\ref{int}) 
as the average value of $z_{D,N}(\Omega_d)$ over randomly chosen points 
$\Omega_d$ on the unit sphere $S^d$. Such points can easily be generated
from Gaussian random numbers \cite{recipes}  $x_i, i=1, \ldots, d$, 
projected onto $S^d$.

We now face an a priori formidable technical difficulty:  One needs
an efficient algorithm for evaluating $z_{D,N}(\Omega_d)$.  However,
it is  well-known that Pfaffians belong to the {\it enfants terribles}
of numerical analysis.  Direct evaluation of eq.(\ref{pfaff}) is
impossible. This has to be compared to the case of determinants,
where very powerful algorithms are available. There {\it is } a
direct, well-known relationship between Pfaffians and determinant
since one has $(\Pf\CM)^2=\det\CM$. Taking the square-root, we can
always find the Pfaffian up to a sign. Unfortunately, we have found
that for $N>2$ the Pfaffians\footnote{An exception seems to be 
dimension $D=4$, where we conjecture the Pfaffian $\CP_{4,N}$ to be positive
semi-definite for all $N$.}
are {\it not} positive semi-definite!
As discussed in detail earlier, the
direct computation of the Pfaffian is possible in generality only
for $D=4,6$.  For $D=10$ more ingenuity is required,
except for $SU(2)$, where a convenient representation is available, 
cf eq.(\ref{twopfaff}).  Using $SO(10)$ invariance, we have also
found a way around the obstacle in the case of $D=10$, $SU(3)$:
cf eqs.(\ref{pfafftodet}),(\ref{redint}). $D=10$ and $N \geq 4$
require new ideas, which will not be treated in this paper.

After computation  of $z_{D,N}(\Omega_d)$, the evaluation of $\CZ_{D,N}$
with the  simple ``direct sampling'' algorithm is straightforward, as soon
as  the variance of the integral exists and is sufficiently small
(cf~eq.(\ref{variance})).  This latter condition is far from
innocent, as can be seen from the data in table 1, where we present
our results for $SU(2)$. We actually give the mean in units of
$\sqrt{8 \pi}$, cf eq.(\ref{sutwo}). \\

\begin{center}
Table 1: Data for SU(2), direct evaluation of integral\\
\vspace{0.5cm} 
\begin{tabular}{||r|c|c|c||} \hline
D   & samples  &  Mean  & error       \\ \hline
 4          & $4\times 10^{12}$  & 0.242  & $\infty$ \\
 6          & $2\times 10^{12}$  & 0.250  & $<0.001$  \\ 
10          & $1\times 10^8$     & 1.250  & $<0.001$   \\ \hline
\end{tabular}\\
\end{center} 

Indeed, we have been unable to compute the $D=4$ integral, because
the variance of the integral does not exist in that case. Let us
mention in passing that the (absolute) divergence of any integral
$\int dx f(x)$
is best checked not by direct sampling as in
eqs.(\ref{direct}),(\ref{havelz1}), but by performing a separate Metropolis
random walk, to be discussed below, 
with $\pi = |f|$. For the case of $f = z_{D=4,N=2}^2$,
the Markov chain of this random walk quickly gets stuck in one valley,
which proves the divergence of the {\it second moment} 
for all intents and purposes. For the integral itself the random walk
finds its way back out of the valley (since the integral is finite),
but it cannot be calculated by the present method.
This extremely powerful
method to ascertain the absolute convergence of smooth high-dimensional
integrals plays no role in usual statistical physics applications,
where the weight function $\pi(x)$ (Boltzmann weight or density
matrix for classical and quantum applications, respectively) always
assures convergence. With this simple method, clearcut answers can
be obtained both for the absolute convergence of the integral itself
{\it and} for the variance. In cases of doubt, the convergence
check has to be done first since the relative weight of rare but strong
fluctuations is difficult to assess by direct sampling.

Direct sampling of the integral, while successful for $D=6,10$
for $SU(2)$, as shown, has proven unfeasible for higher gauge
groups.  We thus need to consider importance sampling. The canonical
way to proceed would be to use an analytically tractable approximation
$\pi \sim |z_{D,N}|$ with known integral $V_{\pi}$.  Such functions
are unfortunately yet unknown to us. Short of this, we will present
in the remainder of this paper an efficient algorithm which allows us to
compute the ratio $\CZ_{D^>,N}/\CZ_{D^<,N}$ for any of the pairs 
$D^< < D^>$ of interest to us. The idea of the method is that
for any configuration of matrices $X_1,\ldots,X_{D^>}$ describing 
a ``valley'' configuration with large weight, we will be able to find
subensembles of $D^<$ matrices (among the above) describing a configuration
with an important contribution to $\CZ_{D^<,N}$.  

To implement the idea, we first note that one condition 
in the above eq.(\ref{havelz1}) was  rather too restrictive.
Because of our particular way of compactifying the integral, we may
generate random points $\Omega_{d^<}$ on the surface of the sphere
$S^{d^<}$ not only from projected Gaussian variables in 
${\rm I}\!{\rm R}^{d^<}$,
but also from any other volume isotropic  with respect to 
$S_{d^<}$. Surprisingly, 
one perfectly
correct way 
to generate random $\Omega_{d^<}$ consists in
taking $d^<$ components  of random vectors on 
$S_{d^>}$. Generalizing further, we
write
\begin{equation}
\pi(\Omega_{ X_1,\ldots,X_{D^>}}) = \frac{1}{\nu}\sum_{\{\sigma\}} 
z_{D^<,N}( \Omega_{X_{i_1},\ldots,X_{i_{D^<}}})
\label{fillup}
\end{equation}
here $\sigma = (i_1,\ldots, i_{D^<}) $ denotes a combination of
$D^<$ indices out of $D^>$, and $\{\sigma\}$ the sum over all the
combinations. The total number $\nu$ of possible combinations is
\begin{equation}
\nu={D^> \choose D^<}
\end{equation}
In an effort to be completely 
explicit, we specify that $\Omega_{ X_1,\ldots,X_{D^>}}$ and 
$\Omega_{X_{i_1},\ldots,X_{i_{D^<}}}$ denote points on the surface of
$S_{d^>}$ and $S_{d^<}$, respectively.

The important point is now that
the average value of $\pi(\Omega_{ X_1,\ldots,X_{D^>}})$ is equal to
$\CZ_{D^<,N}$. We have just transformed $\CZ_{D^<,N}$ into an integral of 
the same dimension as $\CZ_{D^>,N}$. The last step of our procedure 
consists in using the
{\it same} set of $ X_1,\ldots,X_{D^>}$ for a simultaneous evaluation
of  both integrands:
\begin{equation}
\frac{\CZ_{D^>,N}}{\CZ_{D^<,N}} = 
\frac { \int {\cal D}\Omega_{d^>} \pi(\Omega_{ X_1,\ldots,X_{D^>}}) 
\times
z_{D^>,N}(\Omega_{ X_1,\ldots,X_{D^>}})/\pi(\Omega_{ X_1,\ldots,X_{D^>}})}
{ \int {\cal D}\Omega_{d^>} \pi(\Omega_{ X_1,\ldots,X_{D^>}}) }
\end{equation}
In practice, we generate points $X_1,\ldots,X_{D^>}$ according to
$\pi(\Omega_{ X_1,\ldots,X_{D^>}}) $ with the Metropolis algorithm:
starting from an initial configuration on the unit sphere
$\Omega_{d^>}=(x_1,\ldots,x_{d^>})$, we propose a simple local change of
two of the coordinates, which conserves the norm of the total
vector, to arrive at $\Omega_{d^>}'$.  The proposed move is accepted with
probability $\min(1,\pi(\Omega_{d^>}')/\pi(\Omega_{d^>}))$ (for an elementary
introduction cf, eg, \cite{werner}). 
We also have to cure the sign problem which is completely marginal 
for $D=4,6,10$.
In fact, since $\pi$ in eq.(\ref{fillup}) is not always positive, 
we write
\begin{equation}
\frac{ \int {\cal D}\Omega_d~\pi \times z/\pi}
{ \int {\cal D}\Omega_d~\pi} = 
\frac{ \int {\cal D}\Omega_d~|\pi| \times z/|\pi|}
{ \int {\cal D}\Omega_d~|\pi|~{\rm sgn}(\pi)}
\end{equation}
$|\pi|$ is now a positive weight, which can be used for Monte Carlo sampling
\cite{blankenbecler}.

As a result, we obtain the ratio between two different partition 
functions.
We now present our results on the ratios of the partition 
functions, both for $SU(2)$, which serves as a test,
and for $SU(3)$. Obtaining these data required a few days on ten 
workstations at one of our institutes (AEI Potsdam).
\begin{center}
Table 2: Data for SU(2), ratio of partition functions\\
\vspace{0.5cm}
\begin{tabular}{||r|c|c|c||} \hline
$D^>/ D^<$  & samples  &  Mean  & error       \\ \hline
 6 / 4      & $6.0 \times 10^7$   &  1.00  & $<$0.01        \\
 10/ 4      & $4.0 \times 10^7$   &  5.00  & 0.01        \\
 10/ 6      & $3.8 \times 10^7$  &  5.01  & 0.01        \\ \hline
\end{tabular}
\end{center}
\begin{center}
Table 3: Data for SU(3), ratio of partition functions\\
\vspace{0.5cm}
\begin{tabular}{||r|c|c|c||} \hline
$D^>/ D^<$  & samples  &  Mean  & error       \\ \hline
 6/  4      & $4.7 \times 10^7$   &  0.99  & 0.01        \\
 10/ 4      & $6.0 \times 10^7$   &  9.95  & 0.09        \\
 10/ 6      & $4.0 \times 10^7$   &  9.98  & 0.02        \\ \hline
\end{tabular}
\end{center}

For $SU(2)$, the exact values for the three ratios 
$6/4$, $10/4$, $10/6$ (cf~eq.(\ref{sutwo})) 
are, respectively, $1,5,5$, which are reproduced.
For $SU(3)$, the numerical computation leads to a novel, 
and very strong result:
Here the ratios in question are to a surprisingly good precision 
$1,10,10$, which beautifully agrees with (\ref{ourconject}) for $N=3$.

Unfortunately, the
price to pay for coming up with a powerful algorithm was that we
had to give up computing the integrals themselves. We have not
attempted to test our conjecture eq.(\ref{groupfactor})
for the group factor $\CF_N$ which appears in the proposed form
(\ref{ourconject}) of the partition functions $\CZ_{D,N}$. A direct 
computation of the integrals themselves belongs to the same class 
of Monte Carlo problems as free energy calculations, which
are notoriously difficult. As mentioned before, 
it seems preferable to find a good  analytical 
approximation to the integrands with known integrals. 
On the other hand, we believe that the difficulties in the 
calculation of the $D=10$ Pfaffians can be overcome,
and that our ratio test of the extended Green-Gutperle conjecture
can be extended to gauge groups with $N \geq 4$. 
In any case, and independent on whether the extended conjecture will turn out
to be correct, we feel we have already demonstrated the usefulness
of numerical calculations in getting reliable information on
these fascinating and intricate supersymmetric models.\\
{}\\
{\bf Note added:} 
We are now able to numerically compute the integrals for $SU(3)$, $ SU(4)$ 
and $SU(5)$ directly \cite{kns2} and to confirm the corrected 
formula eq.(\ref{ourconject}). Meanwhile we have also received
the impressive paper \cite{moore}, where under certain assumptions
(i.e.~contour prescriptions) the result for arbitrary $N$
(but without working out $\CF_N$) is arrived at by analytical methods.

\acknowledgements
W.~K., H.~N. and M.~S. thank, respectively, the AEI Potsdam, 
the LPT-ENS Paris and the LPS-ENS Paris for hospitality. 
This work
was supported in part by the EU under Contract FMRX-CT96-0012.

%\end{multicols}
\end{document}